\documentclass[10pt]{article}
\usepackage{amsfonts, amssymb, amsthm,amsmath}
\usepackage{epsfig}
\usepackage{color}

\newcommand{\be}{\begin{equation}}
\newcommand{\ee}{\end{equation}}
\newcommand{\bea}{\begin{eqnarray}}
\newcommand{\eea}{\end{eqnarray}}

\begin{document}

%%%%%%%%%%%%%%%%%%%%%%%%%%%%%%%%%%%%%%

\begin{center}
{\LARGE{\bf{The Fedosov $*$-product in Mathematica }}}
\end{center}

\begin{center}
\vskip0.25cm

{\bf Jaromir  Tosiek}  \\

\vskip0.25cm
{\em  Institute of Physics}\\ {\em Technical University of Lodz}\\ {\em ul. Wolczanska 219, 93-005 Lodz}\\
{\em Poland}

\vskip0.15cm

e-mail:  tosiek@p.lodz.pl

\vskip0.25cm\centerline{\today}

\end{center}

\begin{abstract}
The computer program `Fecom.nb' implementing the Fedosov $*$-product  in Darboux coordinates is presented. It has been written in Mathematica 6.0 but it can be easily modified  to be run in some earlier version of Mathematica. To optimize computations elements of the Weyl algebra are treated as polynomials. Several procedures which order the terms are included. 
\end{abstract}

\section{Introduction}
A phase space formulation of 
 quantum mechanics was introduced by Moyal \cite{MO49} in the middle of the 20th century. Moyal's way of thinking about quantum world arose from earlier works by Weyl \cite{WY31}, Wigner \cite{WI32} and Groenewold \cite{GW46}.
The phase space quantum mechanics in its original version was restricted to systems with  the
   phase spaces ${\mathbb R}^{2n}.$ It was a complete physical theory i.e.  formulas defining a $*$-product of functions, mean values, eigenstates and describing a  time evolution were given.

As far as we know an extension of the Moyal quantum mechanics on nontrivial phase spaces  containing a quantum state representation and a method of searching for eigenvalues and eigenstates does not exist. The starting point  for constructing a  theory of this kind must be some modification of the  Moyal  $*$-product.

The concept of  a $*$- product on an arbitrary phase space was  
  proposed by Bayen {\it et al} \cite{baf}, \cite{bay}. In this approach quantum theory was considered as a deformation of classical mechanics with the Planck constant $h$ as a deformation parameter. 

According to Bayen {\it et al} the $*$-product of smooth functions $A$ and $B$ defined on a phase space can be written as follows
\be
\label{poczatek}
A * B = A \cdot B + \frac{i \lambda}{2} \{A ,B \}_P + \sum_{k=2}^{+ \infty} {\lambda}^{k}  D_k(A,B),
\ee
where $\lambda$ is a deformation parameter, $\{\cdot,\cdot\}_P$ denotes the Poisson bracket and $D_k$'s, $2 \leq k $ are some bidifferential operators. Works by
 Bayen {\it et al}  established only general frames of deformation quantization and did not contain   explicit formulas describing operators $D_k$ so each case required individual analysis.

In 90's of the previous century two outstanding papers devoted to a practical realisation of the  Bayen {\it et al}  quantization program on a symplectic manifold endowed with a symplectic torsionfree connection were published by Fedosov \cite{6}, \cite{7}. He found a purely algebraic construction of the $*$-product determined by this symplectic connection. Since every symplectic manifold may be equipped with a symplectic connection, his method can be applied  on any symplectic manifold. 

Explicit form  of the $*$-product in this formalism is  extremely complicated. However the  Fedosov algorithm is based on two recurrence relations and therefore the $*$-product of two functions $A$ and $B$ can be found up to an arbitrary fixed power of the deformation parameter after a finite number of steps. Moreover, 
his  $*$-product can be computed by a computer program. In this paper we present a  program written in Mathematica 6. To improve efficiency of the code we restrict ourselves to Darboux coordinates. 

Our contribution is a `kernel' on  which procedures for individual purposes can be built.  In the 2nd section of this paper we show, how elements of the Fedosov quantization program have been implemented in the program. The next section is a short instruction for users. Some tests end the paper.    

In all  formulas, in which summation limits are obvious, we use the Einstein summation convention.

Names of procedures  are written in a {\bf bold} typestyle. Names of variables are in {\it italic}. 
%%%%%%%%%%%%%%%%%%%%%%%%%%%%%%%%%%%%%%%%%%%%%%%%%%%%%%%%%%%%%%%%%%%%%%%%%%%%%%%%%%%%%%%%%%%%%%%%%%%%%%%%%%%%%%%%%%%%%%%%%
%%%%%%%%%%%%%%%%%%%%%%%%%%%%%%%%%%%%%%%%%%%%%%%%%%%%%%%%%%%%%%%%%%%%%%%%%%%%%%%%%%%%%%%%%%%%%%%%%%%%%%%%%%%%%%%%%%%%%%%%%
%%%%%%%%%%%%%%%%%%%%%%%%%%%%%%%%%%%%%%%%%%%%%%%%%%%%%%%%%%%%%%%%%%%%%%%%%%%%%%%%%%%%%%%%%%%%%%%%%%%%%%%%%%%%%%%%%%%%%%%%%
%%%%%%%%%%%%%%%%%%%%%%%%%%%%%%%%%%%%%%%%%%%%%%%%%%%%%%%%%%%%%%%%%%%%%%%%%%%%%%%%%%%%%%%%%%%%%%%%%%%%%%%%%%%%%%%%%%%%%%%%%
%%%%%%%%%%%%%%%%%%%%%%%%%%%%%%%%%%%%%%%%%%%%%%%%%%%%%%%%%%%%%%%%%%%%%%%%%%%%%%%%%%%%%%%%%%%%%%%%%%%%%%%%%%%%%%%%%%%%%%%%%
%%%%%%%%%%%%%%%%%%%%%%%%%%%%%%%%%%%%%%%%%%%%%%%%%%%%%%%%%%%%%%%%%%%%%%%%%%%%%%%%%%%%%%%%%%%%%%%%%%%%%%%%%%%%%%%%%%%%%%%%% 

\section{  The Fedosov's construction of deformation quantization}

\setcounter{pr}{0}
\setcounter{co}{0}
\setcounter{tw}{0}
\setcounter{de}{0}
\setcounter{equation}{0}

 At the beginning we explain how the Fedosov algorithm has been implemented in our computer program. We assume that the Reader is familiar with the Fedosov method. For details see  \cite{7}.

Let $({\cal M}, \omega, \Gamma)$ be a $2n$-D Fedosov manifold and ${\cal A}= \{({\cal U}_z,\phi_z)\}_{z \in J}$  a Darboux atlas on this manifold. By $\omega$ we mean the symplectic $2$-form. In each chart belonging to the atlas ${\cal A}$ 
\be
\label{1}
\omega= \sum_{i=1}^n dx^{i} \wedge dx^{i+n}.
\ee
The symbol $\Gamma$ denotes a symplectic connection i.e. torsionfree connection preserving the symplectic $2$-form. Locally in  Darboux coordinates the symplectic connection is determined by a set of coefficients $\{\Gamma_{ijk} \stackrel{\rm def.}{=} \Gamma^l_{jk}\omega_{il}\}$ symmetric in all indices \cite{6,7,gel}.

Let $h$ denote a deformation   parameter. We assume that it is positive. By $y^1, \ldots, y^{2n}$ we denote the components of an arbitrary vector ${\bf y}$ belonging to the tangent space $T_{\tt p} {\cal M}$
 at the point ${\tt p} \in {\cal M} $ with respect to   the natural basis $\left( \frac{\partial}{\partial x^i}\right)_{\tt p}$ determined by the chart $({\cal U}_z,\phi_z)$ such that ${\tt p} \in {\cal U}_z.$

We introduce the formal series 
\be
\label{2}
a \stackrel{\rm def.}{=}\sum_{l=0}^{\infty}h^k a_{k, j_1 \ldots j_l} y^{j_1} \ldots y^{j_l}, \;\; k \geq 0
\ee
at ${\tt p}.$
For $l=0$ we put $a = h^k a_k.$ By $a_{k, j_1 \ldots j_l}$ we denote the components of a covariant tensor symmetric with respect to the indices $\{j_1, \ldots, j_{l}\}$ taken in the basis $dx^{j_1} \odot \ldots \odot dx^{j_l}. $

In our program the deformation parameter $h$ is denoted by h. The components $y^1, \ldots, y^{2n}$ are elements of a vector Yy, so $y^i$  in the computer program is represented by Yy[[i]]=y[i].

The part of the series $a$ standing at $h^k$ and containing $l$ components of the vector ${\bf y}$ will be denoted by
$a[k,l].$ Thus 
\be
\label{2a}
a= \sum_{k=0}^{\infty}\sum_{l=0}^{\infty}h^k a[k,l].
\ee
 The  degree $\deg(a[k,l]) $ of the component $a[k,l]$ is equal to $2k+l.$

Notice that since $a_{k, j_1 \ldots j_l}$ are completely symmetric in the indices $\{j_1, \ldots, j_l\},$ the element $a$ defined by the formula (\ref{2}) can be understood as the polynomial 
\be
\label{3}
a= \sum_{z=0}^{\infty} \sum_{k=0}^{\left[\frac{z}{2}\right]} h^k \tilde{a}_{k,i_1 \ldots i_{2n}}(y^{1})^{i_1} \ldots 
(y^{2n})^{i_{2n}}, 
\ee
where
\[
0 \leq i_1, \ldots, i_{2n}\leq z-2k \;\;\; , \;\;\;
i_1+ \cdots + i_{2n}= z-2k.
\]
The symbol $\left[\frac{z}{2}\right]$ denotes the floor of $\frac{z}{2}.$
The relation between the tensor components $a_{k, j_1 \ldots j_l}$ and the polynomial coefficients $\tilde{a}_{k,i_1 \ldots i_{2n}}$
reads
\be
\label{4}
\tilde{a}_{k,i_1 \ldots i_{2n}}= \frac{(z-2k)!}{i_1! \cdots i_{2n}!}\, a_{k\, {\scriptsize \underbrace{11 \ldots 1}_{i_1\;\;{\rm indices}} \ldots 
\underbrace{2n \, 2n  \ldots 2n}_{i_{2n} \;\;{\rm indices}} }  }.
\ee
Using the polynomials instead of the completely symmetric tensors shortens drastically the time required for computations.

Let $P^*_{\tt p}{\cal M}[[h]]$ denote a set of all elements $a$ of the form (\ref{2}) at the point ${\tt p}. $    
The product $\circ: P^*_{\tt p}{\cal M}[[h]] \times P^*_{\tt p}{\cal M}[[h]]\rightarrow P^*_{\tt p}{\cal M}[[h]]$ of   two elements $
a, b \in P^*_{\tt p}{\cal M}[[h]]$ is the mapping
\be
\label{5}
a \circ b
\stackrel{\rm def.}{=}  \sum_{t=0}^{\infty} \frac{1}{t!}\left(-\frac{ih}{2}\right)^t\omega^{i_1 j_1} \cdots \omega^{i_t j_t} \:\frac{\partial^{ t}  a }{\partial y^{i_1}\ldots\partial y^{i_t} }\:\frac{\partial^{t}  b }{\partial y^{j_1}\ldots\partial y^{j_t} }, 
\ee
where $\omega^{ji}$ is the inverse tensor to $\omega_{kj}$ i.e.
\[
 \omega_{kj} \omega^{ji}= \delta^i_k.
\]
The pair $(P^*_{\tt p}{\cal M}[[h]],\circ) $ is a noncommutative associative algebra called the  Weyl algebra. 
All  $a, b \: \in \:(P^*_{\tt p}{\cal M}[[h]],\circ)$ fulfill the relation 
\be
\label{6}
 \deg(a \circ b)= \deg(a) + \deg(b).
\ee
Our program works in a Darboux chart. In this chart formulas defining the $\circ$-product are simple and we may treat elements of the Weyl algebra as the polynomials instead of the tensors. Moreover, the symplectic connection is represented by an element of the Weyl algebra (see a relation (\ref{15})) and we calculate exterior covariant derivatives in terms of the Weyl algebra (\ref{14}). The $\circ$-product is computed in three steps.
\begin{enumerate}
\item
The module {\bf Circ1} finds the $\circ$-product of the monomials $(y^i)^{r}(y^{n+i})^j \circ (y^i)^{s}(y^{n+i})^k,\\ 1 \leq i \leq n, \;\; 0 \leq r,j,s,k.  $ As it has been proved \cite{ja4}
\[
(y^i)^r (y^{i+n})^j \circ (y^i)^s (y^{i+n})^k
= r!\: j!\: s!\: k!\: \sum_{t=0}^{{\rm min}[r,k]+{\rm min}[j,s]} \left( \frac{i h}{2}\right)^t (y^i)^{r+s-t}(y^{i+n})^{k+j-t}
\times
\]
\be
\label{7}
\times \sum_{a={\rm max}[t-r,t-k,0]}^{{\rm min}[j,s,t]}(-1)^a \frac{1}{ a!\: (t-a)!\:(r-t+a)!\:(j-a)!\:(s-a)!\:(k-t+a)!}.
\ee
\item
The $\circ$-product of the monomials
\[
\Big( (y^1)^{i_1}(y^2)^{i_2}\ldots (y^{2n})^{i_{2n}} \Big) \circ \Big( (y^1)^{j_1}(y^2)^{j_2}\ldots (y^{2n})^{j_{2n}}\Big)=
\] 
\be
\label{8}
\Big( (y^1)^{i_1} (y^{n+1})^{i_{n+1}} \circ (y^1)^{j_1} (y^{n+1})^{j_{n+1}}  \Big) \cdots 
\Big( (y^n)^{i_n} (y^{2n})^{i_{2n}} \circ (y^n)^{j_n} (y^{2n})^{j_{2n}}  \Big)
\ee
is generated in the procedure {\bf Circ2}.
\item
Using bilinearity of the $\circ$-product the module {\bf Circproduct} finds the product of the two polynomials from the Weyl algebra $(P^*_{\tt p}{\cal M}[[h]],\circ) $. 
\end{enumerate}
In the operation of the $\circ$-multiplication of the polynomials $a \circ b$ most of time is used to compute the product (\ref{8}) of the monomials. Hence we prepared the procedure {\bf Ord}, which orders the polynomial in $y^1, \ldots, y^{2n}$ in such a way that every term $(y^1)^{i_1}(y^2)^{i_2}\ldots (y^{2n})^{i_{2n}}$ appears only once. For example
\[
{\rm \bf Ord}\Big[a y[1] + f[x[1],x[2]]y[1]+\frac{1}{2}y[2] + \frac{1}{3}y[1] \Big]= \Big(\frac{1}{3} +a+ f[x[1],x[2]]  \Big)y[1]
+\frac{y[2]}{2}.
\]

Taking a set of the Weyl algebras $(P^*_{\tt p}{\cal M}[[h]],\circ) $ at all  points of the manifold ${\cal M}$ we constitute the Weyl bundle 
\[
{\cal P^*M}[[h]] \stackrel{\rm def.}{=} \bigcup_{{\tt p} \in {\cal M}}   (P^*_{\tt p}{\cal M}[[h]],\circ).
\]

Geometrical structure of the Fedosov deformation quantization is based on an 
 $m$-differential form calculus with values in the Weyl bundle. Locally such a form can be written as follows 
\be
\label{9}
a= \sum_{l=0}^{\infty}h^k a_{k, j_1 \ldots j_l, s_1 \ldots s_m}(x^1, \ldots, x^{2n}) y^{j_1} \ldots  y^{j_l}dx^{s_1} \wedge
\cdots \wedge dx^{s_m},
\ee
where $0 \leq m \leq 2n.$ Now
$a_{k, i_1 \ldots i_l, j_1 \ldots j_m}(x^1, \ldots, x^{2n}) $ are components of smooth tensor fields on ${\cal M}$ and 
$ C^{\infty}({\cal TM})\ni {\bf y}\stackrel{\rm locally }{=} y^i \frac{\partial}{\partial x^i} $ is a smooth vector field  on      
${\cal M}.$ We use the same symbol for the vector field 
$  {\bf y} \in  C^{\infty}({\cal TM})$ and the vector ${\bf y} \in  T_{\tt p}{\cal M}.$ In the program   we treat these objects  as entries of the same vector Yy.

From now on  we will omit the variables $(x^1, \ldots, x^{2n})$ in $a_{k, j_1 \ldots j_l, s_1 \ldots s_m}(x^1, \ldots, x^{2n})$. 

 The forms  (\ref{9}) are smooth sections of the direct sum
$ {\cal P^*M}[[h]] \otimes \Lambda \stackrel{\rm def.}{=} \oplus_{m=0}^{2n}({\cal P^*M}[[h]] \otimes \Lambda^m)
$. By $\Lambda^m$ we mean  a smooth field of $m$-forms on the  manifold ${\cal M}.$
The $\circ$-product $a \circ b$ of two forms $a \in C^{\infty}({\cal P^*M}[[h]] \otimes \Lambda^{m_1})$ and  
$b \in C^{\infty}({\cal P^*M}[[h]] \otimes \Lambda^{m_2})$ is understood as the  $\circ$-product of the Weyl algebra elements and the exterior product `$\wedge$' of the forms.

Computations in  the Fedosov deformation quantization are done on $0$-forms and $1$-forms. Hence we do not prepare a general module calculating the exterior product of forms. We give only  the $\wedge$-multiplication of $1$-forms $dx^i \wedge dx^j, \;   1\leq i,j\leq 2n.$ 

In our program the
elements of a local basis of $\Lambda^1$ are entries of a vector dxx written as dx[i]. 
The product $dx^i \wedge dx^j$ has been denoted as dxx[[i]]**dxx[[j]]. The symbol `**' is used in Mathematica for  some associative but noncommutative product. 

For our purposes it is sufficient to implement antisymmetry of the $\wedge$-product.
Hence we declare  values of the products $dx^i \wedge dx^j$ for $1 \leq i < j \leq 2n$ first and then introduce antisymmetry. The exterior product  ${\rm dxx[[i]]} \wedge  {\rm dxx[[j]]}$ is denoted as $ {\rm {\bf Wedgeproduct}[\,dxx[[i]],dxx[[j]]\,]}$, where 
\[
\forall_{1 \leq i < j \leq 2n} \;
{\rm {\bf Wedgeproduct}[\, dxx[[i]],dxx[[j]]\,]}\stackrel{\rm def.}{=} {\rm dxx[[i]]**dxx[[j]]}
\]
and
\[
{\rm {\bf Wedgeproduct}[x19_{-} ,x19_{-} ]:=0\;\;\;,\;\;\;  {\bf Wedgeproduct}[x13_{-} ,y13_{-} ]:=- {\bf Wedgeproduct}[y13,x13]}.
\]
Therefore for example
 \[
{\rm {\bf Wedgeproduct}[\, dxx[[1]],dxx[[2]]\,]}={\rm dx[1]**dx[2]}, 
\]
\[
{\rm {\bf Wedgeproduct}[\, dxx[[1]],dxx[[1]]\,]}=0,
\]
\[
{\rm {\bf Wedgeproduct}[\, dxx[[2]],dxx[[1]]\,]}= -{\rm dx[1]**dx[2]} .
\]
The $\circ$-product of $1$-forms $a \in {\cal P^*M}[[h]] \otimes \Lambda^1$ and $b \in {\cal P^*M}[[h]] \otimes \Lambda^1$ with values in the Weyl algebra is computed by the module {\bf Fullproduct}.

An important role in our considerations is played by a commutator of forms. By definition 
the commutator of  forms $a \in C^{\infty}({\cal P^*M}[[h]] \otimes \Lambda^{m_1})$ and  $b \in C^{\infty}({\cal P^*M}[[h]] \otimes \Lambda^{m_2})$ is the form $[a,b] \in C^{\infty}({\cal P^*M}[[h]] \otimes \Lambda^{m_1+m_2})$ 
\be
\label{10}
[a,b] \stackrel{\rm def.}{=} a \circ b - (-1)^{m_1 \cdot m_2}b \circ a.
\ee

In the program
the commutator of  $1$-forms $a $ and  $b$  is calculated by the module {\bf Commutator11}.

The operator $\delta^{-1}:C^{\infty}({\cal P^*M}[[h]] \otimes \Lambda^m) \rightarrow C^{\infty}({\cal
P^*M}[[h]]
\otimes \Lambda^{m-1})$ is defined by
\be
\label{11}
\delta^{-1} a = \left\{ \begin{array}{ccl}
&\frac{1}{l+m}\: y^k \frac{\partial }{\partial x^k}\rfloor a \qquad  &{\rm for} \;\;\; l+m>0,   \\[0.35cm]
 & 0 \qquad &{\rm for} \;\;\; l+m=0,
\end{array}\right.
\ee
where $l$ is  the degree of $a$ in $y^j$'s i.e. the number of $y^j$'s. 
The operator $\delta^{-1}$ raises the degree of the forms of ${\cal P^*M}[[h]] \Lambda$ in the Weyl algebra by 
$1$.

The operator $\delta^{-1}$ is linear so we need to implement its action on the monomials $(y^{1})^{i_1} \ldots 
(y^{2n})^{i_{2n}}dx^j $ and $(y^{1})^{i_1} \ldots (y^{2n})^{i_{2n}}dx^j \wedge dx^s. $ By direct calculations it can be checked that
\be
\label{12}
\delta^{-1} \Big( (y^{1})^{i_1} \ldots 
(y^{2n})^{i_{2n}}dx^j 
\Big)= \frac{1}{i_1+ \ldots +i_{2n}+1} \Big(
(y^{1})^{i_1} \ldots (y^{2n})^{i_{2n}}y^j
\Big)
\ee
and
\be
\label{13}
\delta^{-1} \Big( 
(y^{1})^{i_1} \ldots (y^{2n})^{i_{2n}}dx^j \wedge dx^s
\Big)=
\frac{1}{i_1+ \ldots +i_{2n}+2} \Big(
(y^{1})^{i_1} \ldots (y^{2n})^{i_{2n}}y^j  dx^s -
(y^{1})^{i_1} \ldots (y^{2n})^{i_{2n}}y^s  dx^j \Big). 
\ee
Acting of the operator $\delta^{-1}$ on $2$-forms is realized by the  module {\bf Delta1} and on $1$-forms by {\bf Delta0}.
Before using these modules the forms are ordered by {\bf Ordnung2 } and {\bf Ordnung1} respectively so each element
 $(y^{1})^{i_1} \ldots (y^{2n})^{i_{2n}}dx^j \wedge dx^s $ or $(y^{1})^{i_1} \ldots 
(y^{2n})^{i_{2n}}dx^j $  appears only once.

The exterior covariant derivative  $\partial_{\gamma}$ of a form $ a \in C^{\infty}({\cal P^*M}[[h]] \otimes \Lambda^m )$ determined by a connection $1$-form 
$\gamma \in C^{\infty}({\cal P^*M}[[h]] \otimes \Lambda^1)$ is the linear operator 
$
\partial_{\gamma} : C^{\infty}({\cal P^*M}[[h]] \otimes \Lambda^m ) \rightarrow C^{\infty}({\cal P^*M}[[h]] \otimes \Lambda^{m+1} ) 
$
defined in a Darboux chart by the formula
\be
\label{14}
\partial_{\gamma} a \stackrel{\rm def.}{=}da + \frac{i}{ h}[\gamma,a].
\ee

In the case when $\gamma$ represents the induced symplectic connection we put
\be
\label{15}
\gamma \stackrel{\rm denoted}{=}\Gamma= \frac{1}{2}\Gamma_{ijk}y^iy^j dx^k.
\ee 
The $1$-form of the symplectic connection $\Gamma$ is called {\it symplecticconnection1form} and it is computed by a command of the same name.

 The curvature form $R_{\gamma}$ of a connection $1$-form $\gamma$
in a Darboux chart can be expressed by the formula
\be
\label{16}
R_{\gamma}= d \gamma + \frac{i}{2  h}[\gamma, \gamma]=  d \gamma + \frac{i}{  h}\gamma \circ  \gamma. 
\ee

In the program a curvature of some connection {\it kona} is computed by the function {\bf Curvaturecalculation}. Hence the symplectic curvature $2$-form is found as {\bf Curvaturecalculation}[symplecticconnection1form].

The crucial role in the Fedosov deformation quantization is played by an  Abelian connection $\tilde{\Gamma}.$ 
The Abelian connection proposed by Fedosov is of the form
\be
\label{17}
\tilde{\Gamma}= \omega_{ij}y^i dx^j + \Gamma + r,
\ee
where the series $r$ is at least of the degree  $3$ and it is defined by the recurrence relation 
\be
\label{18}
r = \delta^{-1} R_{\Gamma} + \delta^{-1} \left( \partial_{\Gamma} r + \frac{i}{ h}r \circ r \right).
\ee
Let $r[z]$  denote the  component $r[z]\stackrel{\rm def.}{=}\sum_{k=0}^{[\frac{z-1}{4}]}h^{2k}r_{m}[2k,z-4k]dx^m, \;\;z \geq 3$   of $r$   of the degree $z.$ 
As it was shown
 \cite{olga}, \cite{vais2}
\[
r[3]= \delta^{-1}R_{\Gamma},
\]
\be
\label{19}
r[z]=\delta^{-1}\left( \partial_{\Gamma}r[z-1]+ \frac{i}{ h}\sum_{j=3}^{z-2}r[j] \circ r[z+1-j]\right), \;\; z  \geq 4.
\ee 
The Abelian connection is written as the vector {\it matrixAbelconnection}. The element of the degree $z$ is saved as {\it matrixAbelconnection[[z]]}. There is only one exception. Since the term $\omega_{ij}y^i dx^j$ does not appear in further operations, we put  
${\it matrixAbelconnection[[1]]}=0.$ The quantities {\it matrixAbelconnection[[z]]} for $3 \leq z$ are calculated by the function {\bf CorrAbelconnection} and ordered by the procedure {\bf Ordnung1} before saving. The Abelian connection in the form of series  is generated by the procedure {\bf Abelianconnectionseries}.

The subalgebra ${\cal P^*M}[[h]]_{\tilde{\Gamma}}  \subset C^{\infty}( {\cal P^*M}[[h]] \otimes \Lambda^0 )$ consists of flat sections i.e. such that
$
\partial_{\tilde{\Gamma}}a=0.
$
For any $a_0 \in C^{\infty}(\cal{M}) $ there exists a unique section $a \in {\cal P^*M}[[h]]_{\tilde{\Gamma}} $ such that the projection  
\[
 \sigma(a)\stackrel{\rm def.}{=}a|_{{\bf y}=0}=a_0.
\] 
The element $a= \sigma^{-1}(a_0)$ can be found by the iteration
\be
\label{20}
a= a_0 + \delta^{-1} \left( \partial_{\Gamma}a + \frac{i}{ h}[r,a]\right).
\ee
This relation means that
\[
a[0]=a_0,
\]
\be
\label{21}
a[z]=\delta^{-1} \Big( \partial_{\Gamma}a[z-1]+\frac{i}{ h} \sum_{l=1}^{z-2} \Big[r[z+1-l],a[l]\Big]\Big), \;\;\;   z \geq 1.
\ee
Remember that $r[1]=r[2]=0.$
The series representing multiplied functions $A$ and $B$ are written as the vectors {\it matrixA}, {\it matrixB} respectively. The part of degree $z$ is on the $(z+1)$'th position of the vector. The term $a[z]$ is generated by the module {\bf Flatcorrection} and it is ordered by the function {\bf Ord}.

Using the one-to-one correspondence between ${\cal P^*M}[[h]]_{\tilde{\Gamma}}$ and $C^{\infty}(\cal{M})$ we introduce the associative star product `$*$' of functions $a_0,b_0 \in C^{\infty}(\cal{M})$ 
\be
\label{22}
a_0 * b_0 \stackrel{\rm def.}{=} \sigma \Big( \sigma^{-1}(a_0) \circ  \sigma^{-1}(b_0) \Big).
\ee
The $*$- product (\ref{22}) with $\lambda=-h$ is of the form (\ref{poczatek}) and in physics it is interpreted as the quantum multiplication of observables.
Notice that in the Fedosov notation the correspondence principle reads
\[
\{a_0,b_0\}_{P} \rightarrow - \frac{1}{i \hbar}\Big(a_0 *b_0 - b_0 *a_0 \Big).
\] 

There is no need  to calculate the complete  product $\sigma^{-1}(a_0) \circ  \sigma^{-1}(b_0)$ and then to 
project it on ${\cal M}$.
The part of the product $a_0 * b_0$ standing at $h^k$ must be the projection of $\sum_{l=0}^{2k}\sigma^{-1}(a_0)[l]\circ \sigma^{-1}(b_0)[2k-l]$. Since we are looking for the term at  $h^k$ we consider only elements of this form. Moreover,  for $l>0 $ each expression $\sigma^{-1}(a_0)[l]$ or  $\sigma^{-1}(b_0)[2k-l]$ contains at least one $y$ so we deduce that 
\[
a_0 * b_0=a_0 \cdot b_0 +\sum_{k=1}^{\infty}  \sigma \left( \sum_{l=1}^{2k-1} \sigma^{-1}(a_0)[l]\circ \sigma^{-1}(b_0)[2k-l] \right).
\]
Among the products $\sigma^{-1}(a_0)[l]\circ \sigma^{-1}(b_0)[2k-l] $ only these which do not generate any $y$'s are important. Therefore,  for a fixed $l,$ we check which of the series: $\sigma^{-1}(a_0)[l]$ or $\sigma^{-1}(b_0)[2k-l]$ is shorter. Then for each element of this shorter series we look for its counterpart giving in the $\circ$-product a term containing $h^k$ and then multiply these elements.   

Let the term 
\[
 A= h^s \cdot A(x[1], \ldots, x[2n] ) \;(y^1)^{i_1} (y^2)^{i_2}\ldots (y^{2n})^{i_{2n}}
\] 
from $\sigma^{-1}(a_0)[2s+i_1 + \ldots +i_{2n}]$ be chosen.
Its counterpart 
\[
B= h^{k-s-i_1- \ldots - i_{2n}} \cdot B(x[1], \ldots, x[2n] ) \;(y^1)^{i_{n+1}} (y^2)^{i_{n+2}}\ldots (y^{2n})^{i_{n}}
\]
belongs to $\sigma^{-1}(b_0)[2k-2s-i_1- \ldots - i_{2n}].$
Then using (\ref{7}) and (\ref{8}) one gets
\[
\sigma(A \circ B)= h^{k} \cdot A(x[1], \ldots, x[2n] ) \cdot B(x[1], \ldots, x[2n] ) 
(-1)^{i_{n+1}+ \ldots + i_{2n}}
\left(\frac{i}{2} \right)^{i_1+ \ldots +i_{2n}}i_1 ! \cdots i_{2n}!\; .
\]

The final product is represented by the vector {\it productmatrix}. The terms containing $h^k$ are on the $(k+1)$'th position. These terms are computed by the module {\bf Rdegreecorrection}. Appriopriate elements are found by the function {\bf Lookingfor} and the reduced $\circ$-product is calculated by {\bf Partialcirc}.
%%%%%%%%%%%%%%%%%%%%%%%%%%%%%%%%%%%%%%%%%%%%%%%%%%%%%%%%%%%%%%%%%%%%%%%%%%%%%%%%%%%%%%%%%%%%%%%%%%%%%%%%%%%%%%%%%%%%%%%%%
%%%%%%%%%%%%%%%%%%%%%%%%%%%%%%%%%%%%%%%%%%%%%%%%%%%%%%%%%%%%%%%%%%%%%%%%%%%%%%%%%%%%%%%%%%%%%%%%%%%%%%%%%%%%%%%%%%%%%%%%%

\section{  The program description}

\setcounter{pr}{0}
\setcounter{co}{0}
\setcounter{tw}{0}
\setcounter{de}{0}
\setcounter{equation}{0}

After starting the program the User is asked to give $n$ i. e.  one half of the dimension of the phase space $({\cal M}, \omega). $ The number $n$ is usually  the dimension of the configuration space. Then the program reads coefficients of the symplectic connection $\Gamma.$ 
The symbol Gamma[i,j,k] denotes the coefficient $\Gamma_{ijk}.$
Since $\Gamma_{ijk}$ are symmetric in all indices, the computer asks only for the components $\Gamma_{ijk}$ such that $1 \leq i \leq j \leq k \leq 2n.$ 
In the program local coordinates on ${\cal M}$ are elements of the vector {\it xx}. The i-th coordinate is then {\it xx[[i]]} but,
since by the definition {\it xx} = Array[x, 2n], the shorter form
 {\it x[i]}, $i=1, \ldots, 2n$ is acceptable.

In the next step it is required to declare the range of approximation of the $*$- product. In the case when the multiplied functions do not contain the deformation parameter $h, $ the range is just a maximal power of $h$ appearing in the $*$- product. This quantity, denoted by {\it hpower}, determines the number of repetitions in the recurrence procedures (\ref{19}) and (\ref{21}). Indeed, for a given {\it hpower} it is necessary to make $(2\cdot {\it hpower}-1) $ repetitions. 

Finally we declare the functions {\it A} and {\it B} to be multiplied.

 To enable the User following the Fedosov algorithm, the program prints in an expanded form the $1$- form of the symplectic connection $\Gamma,$ its curvature $R_{\Gamma},$ the sum of the symplectic connection and the Abelian correction $r$  and the flat sections  $\sigma^{-1}(A)$ and $\sigma^{-1}(B).$ All these quantities are known to the degree $(2\cdot{\it hpower}-1). $ 

The program do not check if the components of the symplectic connection $\Gamma_{ijk}$ and the multiplied functions $A$ and $B$ are differentiable sufficiently many times. Please, be careful and do not use symbols which may be confused with Mathematica commands.

Some basic information about  variables and modules can be seen after the use of command  ?nameofprocedure.
A table below contains a list of symbols used in our paper and their counterparts in the program.

\begin{tabular}{|c|c|}\hline
Article & program \\ \hline
$h$ & {\it  h}  \\
$x^i$ &{\it  x[i]}  or {\it xx[[i]]}\\
$y^i$ & {\it y[i]}  or {\it Yy[[i]]} \\ 
$dx^i $ & {\it dx[i]}  or {\it dxx[[i]]} \\
$ dx^i \wedge dx^j$ & {\it dx[i] ** dx[j]} or {\it  dxx[[i]] ** dxx[[j]]} \\
$ \Gamma_{ijk}$ & Gamma[i,j,k]  and {\it  symplecticconnection[[i, j, k]]}\\
$\Gamma$ & {\it symplecticconnection1form} \\
$R_{\Gamma}$ & {\it symplecticcurvature}\\
$r[z],\;\; z \geq 3$ & {\it matrixAbelconnection[[z]]}\\
$a[z],b[z], \;\; z \geq 0$ & {\it matrixA[[z+1]], matrixB[[z+1]]} \\
\hline
\end{tabular}

%%%%%%%%%%%%%%%%%%%%%%%%%%%%%%%%%%%%%%%%%%%%%%%%%%%%%%%%%%%%%%%%%%%%%%%%%%%%%%%%%%%%%%%%%%%%%%%%%%%%%%%%%%%%%%%%%%%%%%%%%
%%%%%%%%%%%%%%%%%%%%%%%%%%%%%%%%%%%%%%%%%%%%%%%%%%%%%%%%%%%%%%%%%%%%%%%%%%%%%%%%%%%%%%%%%%%%%%%%%%%%%%%%%%%%%%%%%%%%%%%%%
%%%%%%%%%%%%%%%%%%%%%%%%%%%%%%%%%%%%%%%%%%%%%%%%%%%%%%%%%%%%%%%%%%%%%%%%%%%%%%%%%%%%%%%%%%%%%%%%%%%%%%%%%%%%%%%%%%%%%%%%%
%%%%%%%%%%%%%%%%%%%%%%%%%%%%%%%%%%%%%%%%%%%%%%%%%%%%%%%%%%%%%%%%%%%%%%%%%%%%%%%%%%%%%%%%%%%%%%%%%%%%%%%%%%%%%%%%%%%%%%%%%

\section{  Tests}

\setcounter{pr}{0}
\setcounter{co}{0}
\setcounter{tw}{0}
\setcounter{de}{0}
\setcounter{equation}{0}
 To check efficiency of our program we applied it to calculate  three examples of the Fedosov $*$-product. The computations were done in Mathematica 6.0.1.0 on the platform Microsoft Windows XP Professional x64 version 2003. The computer used the processor AMD Athlon 64x2 Dual 4400+ and it was equipped with 2GB of RAM. During the tests Mathematica was the only one running program.
\begin{enumerate}
\item
The Moyal product for $n=1$.
\newline
This is a  basic example of the Fedosov $*$- product on the flat manifold ${\mathbb R}^2.$
In the Cartesian coordinates  the symplectic connection $1$-form $\Gamma=0.$ Hence the symplectic curvature vanishes and the same happens to the Abelian correction $r.$ We multiplied two functions $w(xx[[1]],xx[[2]]) * v(xx[[1]],xx[[2]])$ up to the terms standing at $h^4.$ Using the function {\bf Timing} we saw that the time used for generation the symplectic connection $1$-form, its curvature and the Abelian correction was negligible. The matrix representing the flat section  $\sigma^{-1}\Big(w(xx[[1]],xx[[2]])\Big)$ was built in 0.219 s and the matrix containing $\sigma^{-1}\Big(v(xx[[1]],xx[[2]])\Big)$ in 0.203 s. The $*$-product was found in 0.016 s.
\item
The general form of the Fedosov $*$-product  on a $4$- dimensional Fedosov space.
\newline
The  expression representing a symplectic connection $1$-form is
\[
\Gamma=\sum_{i,j,k=1}^4 g_{ijk}(x[1],x[2],x[3],x[4])y[i] y[j] dx[k].
\]
The symplectic connection $1$-form was formed by 0.016 s. Its curvature $2$-form was computed in 0.484 s. We searched the $*$-product up to $h^2$ of two functions  $w(x[1],x[2],x[3],x[4])* v(x[1],x[2],x[3],x[4]).$ The  correction $r$ was built in 14 s. The series $\sigma^{-1}\Big(w(x[1],x[2],x[3],x[4])\Big )$ and $\sigma^{-1}\Big(v(x[1],x[2],x[3],x[4])\Big )$ in an expanded form  were generated by approx. 180 s each. Their $*$-product  was found in 0.046 s.
\item
The symplectic space $({\mathbb R}^2, \omega)$ with the symplectic connection determined by coefficients 
\[
\Gamma_{111}= -x[2]\;\; , \;\; \Gamma_{112}= 0 \;\; , \;\;  \Gamma_{122}= 0  \;\; , \;\; \Gamma_{222}= 0
\]
in a global Darboux chart.
\newline
We generated the eigenvalue equation for $x[2]$ up to $h^5.$  This problem was solved in terms of the Fedosov formalism
in \cite{ja2003}. Opposite signs  of $\Gamma_{111}$  in \cite{ja2003} and in our paper were the consequence of 
different sign conventions. The Wigner eigenfunction of $x[2]$ was denoted  by $w[x[1],x[2]].$

The symplectic connection $1$-form $\Gamma= - \frac{1}{2}x[2]  y[1]^2 dx[1] $ and the symplectic curvature
$2$-form \\ $R_{\Gamma}=\frac{1}{2}  y[1]^2 dx[1] \wedge dx[2]$. The time used for finding these two expressions was
negligible.
The  correction $r$ to the Abelian connection was calculated
in 0.016s, the flat sections $\sigma^{-1}(x[2])$ and  $\sigma^{-1}(w[x[1],x[2]])$ appeared after 0.031s and 5.016s
respectively. The final result 
\be
\label{koniec}
x[2] \Big( w[x[1],x[2]] - \frac{h^2}{8}w^{(0,2)}[x[1],x[2]]- \frac{h^4}{128}w^{(0,4)}[x[1],x[2]]  \Big)+\frac{ih}{2}
w^{(1,0)}[x[1],x[2]]
\ee
was presented immediately.\end{enumerate}

{\bf Acknowlegments}

I would like to thank Michal Dobrski  for his help with Mathematica.

%%%%%%%%%%%%%%%%%%%%%%%%%%%%%%%%%%%%%%%%%%%%%%%%%%%%%%%%%%%%%%%%%%%%%%%%%%%%%%%%%%%%%%%%%%%%%%%%%%%%%%%%%%%%%%%%%%%%%%%%%
%%%%%%%%%%%%%%%%%%%%%%%%%%%%%%%%%%%%%%%%%%%%%%%%%%%%%%%%%%%%%%%%%%%%%%%%%%%%%%%%%%%%%%%%%%%%%%%%%%%%%%%%%%%%%%%%%%%%%%%%%
%%%%%%%%%%%%%%%%%%%%%%%%%%%%%%%%%%%%%%%%%%%%%%%%%%%%%%%%%%%%%%%%%%%%%%%%%%%%%%%%%%%%%%%%%%%%%%%%%%%%%%%%%%%%%%%%%%%%%%%%%
%%%%%%%%%%%%%%%%%%%%%%%%%%%%%%%%%%%%%%%%%%%%%%%%%%%%%%%%%%%%%%%%%%%%%%%%%%%%%%%%%%%%%%%%%%%%%%%%%%%%%%%%%%%%%%%%%%%%%%%%%

\end{document}